 \documentclass[10pt]{emulateapj}  

\usepackage{bm}
\usepackage{natbib}
\usepackage{pslatex}
\newcommand{\irm}{{\rm i}}
\renewcommand{\vec}[1]{\bm{#1}}

\usepackage{graphicx}

\begin{document}

\title{Perpendicular Ion Heating By Reduced Magnetohydrodynamic Turbulence}


\author{Qian Xia\altaffilmark{1}, Jean C.\ Perez\altaffilmark{1,2}, Benjamin D.\ G.\ Chandran\altaffilmark{1}, \& Eliot Quataert\altaffilmark{3}}

\altaffiltext{1}{Space Science Center and Department of Physics, University of New Hampshire, Durham, NH; qdy2@unh.edu, benjamin.chandran@unh.edu}

\altaffiltext{2}{Department of Physics, University of Wisconsin at Madison, 1150 University Avenue, Madison, WI 53706, USA; jcperez@wisc.edu}

\altaffiltext{3}{Astronomy Department \& Theoretical Astrophysics Center, 601 Campbell Hall, The University of California, Berkeley, CA 94720, USA; eliot@astro.berkeley.edu}

\begin{abstract}
Recent theoretical studies argue that the rate of stochastic ion
heating in low-frequency Alfv\'en-wave turbulence is given by $Q_\perp
= c_1 [(\delta u)^3 /\rho] \exp(-c_2/\epsilon)$, where $\delta
u$ is the rms turbulent velocity at the scale of the ion
gyroradius~$\rho$, $\epsilon = \delta u/v_{\perp \rm i}$, $v_{\perp \rm i}$ is
the perpendicular ion thermal speed, and $c_1$ and $c_2$ are
dimensionless constants. We test this theoretical result by
numerically simulating test particles interacting with strong reduced
magnetohydrodynamic (RMHD) turbulence. The heating rates in our
simulations are well fit by this formula.  The best-fit values of
$c_1$ are $\sim 1$. The best-fit values of $c_2$ decrease (i.e.,
stochastic heating becomes more effective) as the grid size and
Reynolds number of the RMHD simulations increase.  As an example, in a
$1024^2 \times 256$ RMHD simulation with a dissipation wavenumber of
order the inverse ion gyroradius, we find $c_2 = 0.21$. We show that
stochastic heating is significantly stronger in strong RMHD turbulence
than in a field of randomly phased Alfv\'en waves with the same power
spectrum, because coherent structures in strong RMHD turbulence
increase orbit stochasticity in the regions where ions are heated most
strongly. We find that $c_1$ increases by a factor of~$\sim 3$ while
$c_2$ changes very little as the ion
thermal speed increases from values $\ll v_{\rm A}$ to values $\sim
v_{\rm A}$, where $v_{\rm A}$ is the Alfv\'en speed. We discuss the
importance of these results for perpendicular ion heating in the solar
wind.
\end{abstract}

\maketitle
\vspace{0.2cm}
\section{Introduction}
\label{sec:intro}
\vspace{0.2cm}

Beginning in the 1950s, a number of authors developed
hydrodynamic models of the solar wind with heating from thermal
conduction~\citep{parker58,parker65,roberts72,durney72}.  For
realistic values of the coronal density and temperature, these models
led to wind speeds near Earth of~$\sim 300$~km/s~\citep{durney72},
much smaller than the speeds of $700-800$~km/s that are observed in
the fast solar wind. In addition, in two-fluid (proton plus electron)
models in which thermal conduction is the only heating mechanism, the
proton temperatures near Earth are much smaller than the observed
proton temperatures~\citep{hartle68}. These discrepancies imply that
the fast solar wind is heated by some mechanism(s) other
than thermal conduction. 

Some clues into the nature of this additional heating are provided by
measurements of ion temperatures.  In-situ spacecraft measurements
show that $T_{\perp \rm p} > T_{\parallel \rm p}$ in low-$\beta$
fast-solar-wind streams, where $\beta = 8 \pi p / B^2$ is the ratio of
the plasma pressure to the magnetic pressure, and $T_{\perp \rm p}$
($T_{\parallel \rm p}$) is the perpendicular (parallel) temperature of
the protons, which measures the speed of thermal motions in directions
perpendicular (parallel) to the magnetic
field~\citep{marsch82b}. Remote observations from the Ultraviolet
Coronagraph Spectrometer show that the temperature of heavy ions is
much larger than the proton temperature in coronal holes
\citep{kohl98, esser99}. In addition, $T_\perp \gg T_\parallel$ for
${\rm O}^{+5}$ ions in coronal holes at heliocentric distances~$r$
of~$\sim 2 R_s$, where $R_s$ is the solar radius~\citep{antonucci00}.
These observations show that, when $\beta \ll 1$, ions experience
strong perpendicular heating and heavy ions are heated preferentially.


One possible mechanism for explaining these ion-temperature signatures
is Alfv\'en waves (AWs) or AW
turbulence~\citep{coleman68}. Propagating AWs are seen in infrared
observations of the solar corona~\citep{tomczyk07}.  AW-like motions
are also seen in optical observations of the low corona, and the
speeds of these motions imply an outward AW energy flux sufficient to
power the solar wind~\citep{depontieu07}.  Farther from the Sun,
fluctuations in the magnetic field, electric field, and average proton
velocity are consistent with a broad spectrum of turbulent,
Alfv\'en-wave-like fluctuations, in the sense that there is a rough
equipartition between the fluctuating magnetic energy and kinetic
energy over a broad range of
scales~\citep{tumarsch95,goldstein95a,bale05}.
A hallmark of three dimensional turbulence, in hydrodynamic fluids as well as
magnetized plasmas, is the cascade of fluctuation energy from large scales to
small scales or, equivalently, from small wavevectors~$\bm{k}$ to large
wavevectors~\citep{k41,iroshnikov63,kraichnan65}.  The frequency of a linear AW
is~$k_\parallel v_{\rm A}$, where $k_\parallel$ ($k_\perp$) is the component
of~$\bm{k}$ parallel (perpendicular) to the magnetic field, 
\begin{equation} 
v_{\rm A} = \frac{B}{\sqrt{4 \pi n_{\rm proton} m_{\rm p}}}
\label{eq:defvA} 
\end{equation} 
is the (proton) Alfv\'en speed, $n_{\rm proton}$ is the proton
density, and $B$ is the magnetic field strength. If AW energy cascades
to sufficiently large~$k_\parallel$, then the wave frequency will
become comparable to the cyclotron frequencies of ions, and strong ion
cyclotron heating will result~\citep{ise83,hollweg02}. In principle,
such cyclotron heating could account for the perpendicular ion heating
discussed above.


There is, however, a problem with the above scenario. When turbulent
AWs interact in strongly turbulent plasmas, their energy cascades
primarily to larger~$k_\perp$, and only weakly to
larger~$k_\parallel$~\citep{shebalin83,goldreich95}. As a consequence,
the small-scale waves produced by the anisotropic AW cascade are
unable to cause cyclotron
heating~\citep{quataert98,cranmer03,howes08a}. If the small-scale
waves produced by the cascade were to damp via linear wave damping,
they would lead primarily to parallel electron heating in the
low-$\beta$ conditions of the solar corona, and to virtually no ion
heating.

This discrepancy has led a number of authors to go beyond linear wave
theory in analyzing the dissipation of AW turbulence.
\citep[e.g.,][]{voitenko04,dmitruk04,markovskii06,parashar09,lehe09,lynn12,servidio11a}. In
this paper, we investigate a nonlinear heating mechanism called
``stochastic heating,'' which arises when fluctuating electric and/or
magnetic fields at wavelengths comparable to a particle's gyroradius
disrupt a particle's smooth gyromotion, leading to the
non-conservation of the particle's magnetic moment
\citep{1987PhRvL..59.1436M,johnson01,chen01,chaston04,fiksel09,chandran10b}.
\cite{chandran10} derived an analytic formula for the stochastic ion
heating rate~$Q_{\perp \rm stoch}$ in plasmas with $\beta \lesssim 1$
as a function of the turbulence amplitude at the gyroradius scale.
Their formula (Equation~(\ref{eq:Q2}) below) contains two
dimensionless constants. \cite{chandran10} simulated test-particles
interacting with randomly phased AWs and kinetic Alfv\'en waves~(KAWs)
to evaluate these constants. In this work, we re-evaluate these
constants by simulating test-particles interacting with strong reduced
magnetohydrodynamic (RMHD) turbulence.  The remainder of this paper is
organized as follows. We briefly review previous work on stochastic
ion heating and strong RMHD turbulence in Sections~\ref{sec:stoch}
and~\ref{sec:RMHD}. We then discuss our numerical methods in
Section~\ref{sec:simulation}, present our results in
Section~\ref{sec:results}, and summarize our
conclusions in Section~\ref{sec:cons}.

\vspace{0.2cm}
\section{Stochastic Ion Heating}
\label{sec:stoch}
\vspace{0.2cm}

If an ion moves in the presence of electric and magnetic fields that
vary over a characteristic spatial scale~$l$ and time scale~$\tau$,
and if the ion's gyroradius~$\rho$ and cyclotron frequency~$\Omega_{\rm i}$
satisfy the inequalities~$\rho \ll l$ and $\Omega_{\rm i} \tau \gg 1$, then
the ion's motion in the plane perpendicular to~$\bm{B}$ is nearly
periodic. As a consequence, the ion's magnetic moment~$\mu =
mv_\perp^2/2B$, an adiabatic invariant, is almost exactly conserved
\citep{kruskal62}. Here, $m$ is the ion's mass, $v_\perp$ is the
component of the ion's velocity perpendicular to~$\bm{B}$, $\Omega_{\rm i} = q
B/m c$, $q$ is the ion charge, $c$ is the speed of light, and $\rho =
v_\perp / \Omega_{\rm i}$.  On the other hand, if $l \sim \rho$, and if the
amplitudes of the fluctuations in the electric and/or magnetic fields
are sufficiently large, then the ion's motion in the plane
perpendicular to~$\bm{B}$ ceases to be nearly periodic, even if
$\Omega_{\rm i} \tau \gg 1$, and magnetic moment conservation is violated
\citep{1987PhRvL..59.1436M}. In this case, the velocity-space average of the
ion magnetic moment, $k_{\rm B} T_\perp/B$, can increase in
time. Perpendicular heating resulting from the disruption of particle
gyro-orbits by turbulent fluctuations with $\Omega_{\rm i} \tau $
significantly greater than unity is called stochastic heating.

\cite{chandran10} derived an analytic formula for the stochastic ion
heating rate in anisotropic AW/KAW turbulence, in which the
fluctuating quantities vary rapidly in directions perpendicular to the
background magnetic field and slowly in the direction parallel to the
background magnetic field. The assumptions in their derivation apply
to RMHD turbulence even in the absence of the kinetic physics that
modifies linear waves at lengthscales smaller than the proton
gyroradius~$\rho_{\rm p}$. \cite{chandran10} considered a Maxwellian
distribution of ions with temperature~$T_{\rm i}$, perpendicular
thermal speed
\begin{equation}
v_{\perp \rm i} = \left(\frac{2 k_{\rm B} T_{\rm i}}{m}\right)^{1/2},
\label{eq:vperpi} 
\end{equation} 
and thermal-particle gyroradius
\begin{equation}
\rho_{\rm i} = v_{\perp \rm i}/\Omega_{\rm i}.
\label{eq:rhop} 
\end{equation} 
Using phenomenological arguments, these authors derived the
following analytic formula for the stochastic ion heating rate (per
unit mass) for the case in which $\beta \lesssim 1$:
\begin{equation}
Q_{ \perp \rm stoch} = \frac{c_1(\delta u)^3}{\rho_{\rm i}}
\,\exp\left(-\,\frac{c_2}{\epsilon_{\rm i}}\right),
\label{eq:Q2} 
\end{equation} 
where  $c_1$ and $c_2$ are dimensionless constants,
\begin{equation}
\epsilon_{\rm i} = \frac{\delta u}{v_{\perp \rm i}},
\label{eq:defeps} 
\end{equation} 
\begin{equation}
\delta u = \left[\int_{k_-}^{k_+} E_u(k_\perp) dk_\perp \right]^{1/2},
\label{eq:defvp} 
\end{equation} 
$k_\pm=e^{\pm0.5}/\rho_{\rm i}$ and $E_u(k_\perp)$ is the 1D power
spectrum of the $\bm{E}\times 
\bm{B}$ velocity of the plasma ($c \bm{E} \times \bm{B}/B^2$). The
normalization of $E_{\rm u}(k_\perp) $ is chosen so that
$\int_0^{\infty} E_u(k_\perp) dk_\perp$ is the total mean square
$\bm{E}\times \bm{B}$ velocity.  Thus, $\delta u$ is the rms amplitude
of the $\bm{E}\times \bm{B}$ velocity or ``fluid velocity'' at
scale~$\rho_{\rm i}$.  \cite{chandran10} numerically simulated
stochastic heating of test particles by a spectrum of randomly phased
AWs and KAWs whose spectra are drawn from the critical-balance model
\citep{goldreich95, cho04b}, for the case in which~$\beta \ll 1$.
Their numerically computed heating rates agreed well with
Equation~(\ref{eq:Q2}) with $c_1 = 0.75$ and $c_2 = 0.34$.  However,
they argued that stochastic heating is more efficient at fixed $\delta
u$ in strong AW/KAW turbulence than in a randomly phased wave field,
implying a larger value of~$c_1$ and/or smaller value of~$c_2$,
because strong AW/KAW turbulence produces coherent structures that
increase orbit stochasticity \citep{dmitruk04}. In this paper, we test
this argument and obtain new values of $c_1$ and $c_2$ for the case of
test particles interacting with strong RMHD turbulence.

When the perpendicular length scale $\lambda_\perp$ of a turbulent
``eddy'' or ``wave packet'' is $ \sim \rho_\irm $, the cascade time in
``balanced'' (i.e., zero-cross-helicity) RMHD turbulence is $\tau
_\lambda \sim \rho_{\rm i}/ \delta u$. Thus, $\epsilon_{\rm i} \simeq (\Omega_\irm
\tau _\lambda) ^{-1}$. For critically balanced
turbulence, the linear frequency $\omega_{\rm A} =
k_\parallel v_{\rm A}$ is comparable to the nonlinear
frequency~$\tau_{\lambda}^{-1}$~\citep{goldreich95,
  boldyrev05}. Combining the above relations, we obtain 
\begin{equation}
\omega_{\rm  A} \sim \epsilon_{\rm i} \Omega _\irm.
\label{eq:omegaA} 
\end{equation} 

Equation~(\ref{eq:omegaA}) can be used to estimate the relative
importance of stochastic heating and cyclotron heating for particles
interacting with small-amplitude, randomly phased AWs or KAWs with
$k_\perp \rho_{\rm i} \sim 1$, where the wave frequencies are chosen
to satisfy the critical-balance condition $\omega \sim \rho_{\rm
  i}/\delta u$.  Assuming that the imaginary part of the frequency is
much less than the real part, each ion species makes a contribution
$\gamma_{\rm i}$ to the total wave damping rate that can be expressed
analytically~\citep{kennel67}. If an ion species has an isotropic
Maxwellian distribution with thermal speed~$v_{\perp \rm i}$, then
$\gamma_{\rm i} \propto e^{-\zeta^2}$, where $\zeta = (\omega -
\Omega_{\rm i})/k_\parallel v_{\perp \rm i}$. In writing this
expression, we have retained only the contribution to~$\gamma_{\rm i}$
from the lowest cyclotron harmonic, which is dominant because we
assume $\omega \ll \Omega_{\rm i}$. This exponential factor is
proportional to the number of particles whose parallel velocities
satisfy the wave-particle resonance condition $\omega - k_\parallel
v_\parallel = n\Omega_{\rm i}$ with $n=1$.  Using
Equation~(\ref{eq:omegaA}) and the inequality $\omega \ll \Omega_{\rm
  i}$, we obtain $\zeta^2 \sim 1/(\beta_{\rm i} \epsilon^2)$, where
\begin{equation}
\beta_{\rm i} \equiv \frac{v_{\perp \rm i}^2}{v_{\rm A}^2}.
\label{eq:defbetai} 
\end{equation} 
Because the energy gained by the particles equals the energy lost by
the waves, we can obtain a rough estimate of the ratio of the
cyclotron heating rate $Q_{\rm c}$ to the stochastic heating
rate~$Q_\perp$ by comparing just the
exponential factors in the expressions for $Q_{\rm c}$ and $Q_\perp$,
which yields $Q_{\rm c}/Q_\perp \sim \exp(c_2 \epsilon^{-1}
-\zeta^2)$. This implies that $Q_{\rm c} \ll Q_\perp$ when $\epsilon
\beta_{\rm i} \ll 1$, indicating that cyclotron heating becomes
increasingly subdominant to stochastic heating as $\epsilon$ and/or
$\beta_{\rm i}$ decreases.  To the extent that arguments from linear
wave theory describe cyclotron heating by strong AW/KAW turbulence,
the above discussion also implies that $Q_{\rm c} \ll Q_\perp$ in
strong AW/KAW turbulence when $\epsilon \beta_{\rm i} \ll 1$.

We note that the theory of \cite{chandran10} does not in general apply
to electrons. It is assumed in that theory that the dominant
contribution to the stochastic heating rate comes from fluctuations
with perpendicular lengthscales comparable to the particle gyroradii,
and that these gyroradii are~$\gtrsim \rho_{\rm p}$. In plasmas with
electron temperatures that are comparable to the proton temperature,
the thermal-electron gyroradii are~$\ll \rho_{\rm p}$.  Energetic
electrons could have gyroradii $\gtrsim \rho_{\rm p}$, but the value
of~$\epsilon$ for such electrons would be extremely small, indicating
that stochastic heating of these fast electrons would be exponentially
weak.

\vspace{0.2cm}
\section{Strong RMHD Turbulence}
\label{sec:RMHD}
\vspace{0.2cm}

As described in the introduction, we consider test particles
interacting with strong reduced magnetohydrodynamic (RMHD) turbulence.
The main assumptions of RMHD are that: (1) the fluctuating magnetic
field $\delta \bm{B}$ is much smaller than the background magnetic
field~$\bm{B}_0$; (2) $\delta \bm{B}$ and the fluctuating fluid
velocity~$\delta \bm{u}$ are perpendicular to~$\bm{B}_0$ (i.e., the
fluctuations are ``transverse''); (3) $\nabla \cdot \bm{u} = 0$; (4)
the fluctuations vary much more rapidly in directions perpendicular
to~$\bm{B}_0$ than the direction parallel to~$\bm{B}_0$ ($k_\perp \gg
|k_\parallel|$); (5) the perpendicular lengthscales are much larger
than the proton gyroradius~$\rho_{\rm p}$, and (6) the frequencies of
the fluctuations are much smaller than the proton cyclotron
frequency~\citep{kadomtsev74,strauss76,zank92,schekochihin09}. Although
RMHD is a fluid theory, it is a good approximation for transverse,
low-frequency, non-compressive fluctuations with $k_\perp \gg
|k_\parallel|$ even in collisionless plasmas such as the solar
wind~\citep{schekochihin09}. However, in some cases, such as the solar
wind at $r = 1 \mbox{ AU}$, RMHD is applicable only at lengthscales
that are sufficiently small that $|\delta \bm{B}| \ll B_0$.

In full (as opposed to reduced) magnetohydrodynamics (MHD), there are
three propagating waves: the Alfv\'en wave (AW), the fast magnetosonic
wave, and the slow magnetosonic wave. RMHD retains only one of these
linear wave modes, the~AW. We thus at times refer to RMHD turbulence
as AW turbulence. We recognize that not all types of AW turbulence can
be described within the framework of RMHD. In particular,
Alfv\'en-wave fluctuations with $|k_\parallel| \gtrsim k_\perp$ do not
satisfy the assumptions of RMHD. However, our focus is on strong,
anisotropic AW turbulence, which satisfies $k_\perp \gg |k_\parallel|$
at scales much smaller than the driving
scale~\citep{goldreich95,maron01,perez08a}.

The equations of RMHD can be expressed in terms of the Els\"asser variables
\begin{equation}
\bm{z}^\pm = \bm{u} \mp \bm{b},
\label{eq:zpm} 
\end{equation} 
where $\bm{b} = \delta \bm{B} / \sqrt{4 \pi \rho_0}$ and $\rho_0$ is
the mass density. These equations take the form
 \begin{equation}
        \frac{\partial \bm{z}^\pm}{\partial t}\pm\left(
        \bm{v}_A\cdot \nabla \right)
        \bm{z}^\pm+\left(\bm{z}^\mp \cdot \nabla 
        \right) \bm{z}^\pm = -\nabla P+\nu \nabla^2
        \bm{z}^\pm+ \bm{f}^\pm \label{eq:ch1incompress}
    \end{equation}
\begin{equation}
\nabla \cdot \bm{z}^\pm = 0,
\label{eq:inc} 
\end{equation} 
and
\begin{equation}
\bm{z}^\pm \cdot \bm{B}_0 = 0,
\label{eq:transverse} 
\end{equation} 
where $\bm{v}_A = \bm{B}_0/ \sqrt{4 \pi \rho_0}$ is the
Alfv\'{e}n velocity, $P = (p/\rho_0 + b^2 /2)$, $p$ is the plasma
pressure, $\nu$ is the viscosity (which we have taken to be equal to
the resistivity), and $\bm{f}^\pm$ is an external driving force,
which we include as a source term for turbulence.
Both $\rho_0$ and $\bm{B}_0$ are taken to be constant.

The properties of RMHD turbulence have been studied extensively with
the use of direct numerical
simulations~\citep{dmitruk03b,dmitruk05,perez08a,perez09a,mason12,perez12}. 
These simulations find that AW energy
cascades from large perpendicular scales to small perpendicular
scales, ultimately dissipating at a small scale, which we call the
``Kolmogorov scale,'' in analogy to hydrodynamic turbulence. Scales
much smaller than the energy-injection scale but much larger than the
Kolmogorov scale are referred to as the ``inertial range of scales.''
As in hydrodynamic turbulence, the breadth of the inertial range
increases with increasing Reynolds number $\mbox{Re} = u_{\rm rms}
L_\perp/\nu$, where $u_{\rm rms}$ is the rms amplitude of the
velocity, and $L_\perp$ is the lengthscale characterizing the forcing
term. In ``balanced'' RMHD turbulence, in which there is equal energy
in $z^+$ fluctuations and $z^-$ fluctuations, the inertial-range power
spectrum of the total energy (kinetic plus magnetic) is proportional
to $k_\perp^{-3/2}$ in simulations of strong RMHD
turbulence~\citep{perez08a}. However, the velocity power
spectrum~$E_u(k_\perp)$ is flatter than the total-energy spectrum
(i.e., $\propto k_\perp^{-n}$ with $n< 3/2$), while the magnetic power
spectrum is somewhat steeper than the total-energy
spectrum~\citep{boldyrev11a,boldyrev12}. 





\vspace{0.2cm}
\section{Numerical Method}
\label{sec:simulation}
\vspace{0.2cm}

Our basic numerical method is to solve the RMHD equations in a
periodic, 3D domain using a pseudo-spectral code, and to numerically
integrate the orbits of test particles that propagate within the
time-varying electromagnetic fields produced by the 3D RMHD
simulations. In the following two subsections, we describe the details
of our numerical algorithms.

\vspace{0.2cm}
\subsection{The RMHD Code}
\label{sec:RMHDnum}
\vspace{0.2cm}

To solve the RMHD equations, we use the pseudo-spectral ``RMHD Code''~\citep{perez08a}. The numerical domain of this code is a 3D box with periodic boundary conditions. The background (mean) magnetic field in the simulation, $\bm{B}_0$, is along the $z$ axis, where $(x,y,z)$ are Cartesian coordinates.  We define $k_\perp = \sqrt{k_x^2 + k_y^2}$. That is, we define $k_\perp$ with respect to the mean magnetic field. (In contrast, as discussed below, we define the the test-particle velocity components $v_\perp$ and $v_\parallel$ with respect to the local magnetic field direction.)


The lengths of the box in the $x$, $y$, and $z$ directions are,
respectively, $L_\perp, L_\perp$, and $L_\parallel$, where
$L_\parallel / L_\perp = 6$. The simulations are run for a long
enough time so that the turbulence reaches an approximate statistical 
steady state, and the driving term $\vec f^\pm$ is chosen so that the
rms fluctuating velocity 
\begin{equation}
u_{\rm rms} = \left[\int_0^\infty E_u(k_\perp) dk_\perp\right]^{1/2}
\label{eq:defvrms} 
\end{equation} 
is approximately $v_{\rm  A}/5$. Therefore,
\begin{equation}
\chi = \frac{v_A L_\perp}{u_{\rm rms} L_\parallel } \simeq 1 \label{eq:ch4tim_ratio},
\end{equation}
which means the turbulence is in critical balance at the outer
scale~\citep{goldreich95}. Moreover, as we
discuss further below, the RMHD equations are invariant when
$L_\parallel$ and $v_{\rm A}$ are both multiplied by the same
factor~$\xi$. We take advantage of this fact to enable a single
simulation to appear as different turbulent fields (with different
scaling factors~$\xi$) to different cohorts of test particles.

The driving forces $\bm{f}^\pm$ lie in the $xy$ plane, are
solenoidal and nonzero only at small wavenumbers satisfying
$2\pi/L_\perp \leq k_\perp \leq 4\pi/L_\perp$ and $ 2\pi / L_\parallel
\leq k_\parallel \leq 4\pi / L_\parallel$. We assign random values
(drawn from a Gaussian distribution) to each nonzero Fourier component
of~$\bm{f}^\pm$ at selected times $t_n=nt_{\rm force}$, where
$n=-1,0,1,2,\ldots$, and choose $t_{\rm force}=\tau/5$ so that the
coefficients are refreshed 5 times every eddy turnover time $\tau =
L_\perp/(2\pi\delta u)$. Between these discrete times, such as
$t\in(t_{n+1},t_{n+2})$, we determine the value of $\bm{f}^\pm(t)$ by cubic
interpolation, using the values of $\vec f^\pm(t_n),\vec
f^\pm(t_{n+1}),\vec f^\pm(t_{n+2})$ and $\vec f^\pm(t_{n+3})$. We include $t_{-1}$, 
so that we can begin cubic interpolation at the beginning of the
simulation at~$t=0$ with the desired value for $\vec f^\pm(t=0)$, which
is set to zero in this work. The time
step $\delta t$ of the RMHD simulations is constrained by a Courant
condition to ensure numerical stability and is much smaller than
$t_{\rm force}$.


\vspace{0.2cm}
\subsection{Particle Tracing}
\label{sec:particle}
\vspace{0.2cm}

We introduce test particles into the RMHD simulations after a time
of at least $10 L_\perp/u_{\rm rms}$ has elapsed, so that the turbulence has reached
a statistical steady state in which the rate of viscous dissipation
matches (on average) the rate at which the forcing term adds energy
into the flow.  We neglect Coulomb collisions and track each
particle's velocity~$\bm{v}$ and position~$\bm{x}$ by solving the
equations
 \begin{equation}
   \frac{d\bm{x}}{dt} = \bm{v}  \label{eq:ch4mp}
  \end{equation}
and 
   \begin{equation}
   \frac{d\bm{v}}{dt} = \frac{q}{m}\left( \bm{E} +
   \frac{\bm{v}\times\bm{B}}{c}\right) \label{eq:ch4mv} ,
  \end{equation}
where $\bm{E}$ is the electric field and $\bm{B}$ is
the magnetic field. As when solving the RMHD equations, we use periodic
boundary conditions when tracing particle orbits. If a particle leaves
the simulation domain through one boundary, it re-enters the box from
the corresponding point on the opposite boundary.

In order to simulate perpendicular heating accurately, we want to
minimize the risk that our code artificially violates $\mu$
conservation through numerical error.  \cite{lehe09} showed that a
fourth-order Runge-Kutta integration of Equations~(\ref{eq:ch4mp}) and
(\ref{eq:ch4mv}) leads to a secular decrease in particle energy and
magnetic moment in the presence of a uniform magnetic field and zero electric field. In
particular, when they used ten time steps per gyration, the particle
energy decreased by 1\% per gyration. We thus
reject that method as unsuitable for this problem.  Instead, we follow
\cite{lehe09} and others in using the ``Boris-pusher'' method \citep{boris70},
which differences Equations~(\ref{eq:ch4mp}) and~(\ref{eq:ch4mv}) at
time $t_i = i \Delta t$ ($i = 0,1,2, ...$ ) according to the scheme
    \begin{equation}
    	\frac{\bm{x}_{i+1}-\bm{x}_i}{\Delta t} = \bm{v}_{i+1/2}
    \end{equation}
    and
    \begin{equation}
    	\frac{\bm{v}_{i+1/2}-\bm{v}_{i-1/2}}{\Delta t} = \frac{q}{m} \bm{E}_i + \frac{q}{mc}\frac{\bm{v}_{i+1/2}+\bm{v}_{i-1/2}}{2}\times \bm{B}_i \label{eq:ch4boris2} .
    \end{equation}
    Equation (\ref{eq:ch4boris2}) can be re-written as
    \begin{equation}
        \bm{v}^+ - \bm{v}^- = (\bm{v}^+ + \bm{v}^-) \times \frac{\Delta t q}{2mc}\bm{B}_i \label{eq:ch4boris3}
     \end{equation}
with: $\bm{v}^\pm = \bm{v}_{i\pm1/2} \mp \Delta t q/2m
\bm{E}_i$.  Upon taking the dot product of
Equation~(\ref{eq:ch4boris3}) with $\bm{v}^+ + \bm{v}^-$, one
finds that $|v^+|^2=|v^-|^2$.  Thus,  when $\bm{E} = 0$, the
Boris-pusher method conserves particle energy to machine precision.

We evaluate $\bm{E}$ in the RMHD simulations using
the idealized Ohm's Law,
     \begin{equation}
        \bm{E} = -\frac{\bm{u}}{c} \times \bm{B}.
     \end{equation}
We interpolate the electric and magnetic
fields from the grid points to each particle's position. We follow
\citet{lehe09} in using the triangular-shaped cloud (TSC) method to
interpolate the field information in 4 dimensions (both space and
time). To avoid introducing an artificial component of~$\bm{E}$
parallel to the local magnetic field through the interpolation method,
we follow \cite{lehe09} in replacing the interpolated electric field with
the quantity
       \begin{equation}
        \tilde{\bm{E}} = \overline{\bm{E}} + ( \overline{\bm{E}\cdot \bm{B}} - \overline{\bm{E}}\cdot \overline{\bm{B}}) \frac{\overline{\bm{B}}}{||\overline{\bm{B}}||^2} ,
      \end{equation}
so that $ \tilde{\bm{E}}\cdot \overline{\bm{B}} =
\overline{\bm{E}\cdot \bm{B}} $, where the overlines stand for
the TSC interpolation.  Since $\bm{E}$ is perpendicular
to $\bm{B}$ on the gridpoints of the RMHD simulation, the electric field seen
by the particle is perpendicular to the interpolated magnetic field.

We initialize the particles with random velocities drawn from a
Maxwellian distribution and with positions uniformly distributed
throughout the numerical domain.  Before we introduce the test
particles into a simulation, we calculate the velocity power
spectrum~$E_u(k_\perp)$ and the rms amplitude of the fluctuating
velocity~$u_{\rm rms}$ using a time average of the simulation data
over the time interval $(t_1, t_2)$, where $t_1$ is some time after
the turbulence has reached an approximate statistical steady state,
and $t_2$ is the time at which the particles are introduced, which
satisfies $t_2 > t_1$.  For example, in Simulations D1 through~D5 of
Table~\ref{tab:summary}, $t_1$ = $t_{\rm RMHD} + 14 L_\perp/u_{\rm
  rms}$ and $t_2 = t_{\rm RMHD} + 25 L_\perp/u_{\rm rms}$, where
$t_{\rm RMHD}$ is the beginning of the RMHD simulation.  To determine
$v_{\perp \rm i}$ and~$\Omega_{\rm i}$, we then proceed through the
following steps. First, we pick a value for~$\rho_{\rm i}/L_\perp$,
which fixes the value of~$\delta u/u_{\rm rms}$ through
Equations~(\ref{eq:defvp}) and (\ref{eq:defvrms}). Second, we choose
the value of~$\epsilon$ that we wish to simulate.  The values of
$\epsilon = \delta u /v_{\perp \rm i}$ and $\delta u/u_{\rm rms}$ fix
the value of~$v_{\perp \rm i}/u_{\rm rms}$, while the values of
$v_{\perp \rm i}/u_{\rm rms}$ and $\rho_{\rm i}/L_\perp$ determine the
gyrofrequency $\Omega_{\rm i} = v_{\perp \rm i}/\rho_{\rm i}$ in units
of $u_{\rm rms}/L_\perp$. Third, we choose the value of $\beta_{\rm
  i}$, which is defined in Equation~(\ref{eq:defbetai}), for the
initial particle distribution.  Because $v_{\perp \rm i}/u_{\rm rms}$
is already fixed from the first two steps above, we need to vary
$u_{\rm rms}/v_{\rm A}$ in order to vary~$\beta_{\rm i}$. Here, we
make use of the fact that the RMHD equations are invariant when we
multiply both $v_{\rm A}$ and $L_\parallel$ by the same scaling
factor. For an RMHD simulation with some given value of $v_{\rm
  A}/u_{\rm rms}$ and $L_\parallel/L_\perp$, we re-scale both $v_{\rm
  A}$ and $L_\|$ before passing the field variables to the particle
integrator in order to achieve the desired value of~$\beta_{\rm
  i}$. 

Because we need high accuracy to ensure that the changes of the
particles' magnetic moments are not the result of numerical errors,
the time steps of the test-particle integration are smaller than the
time steps of the RMHD code (by a factor that is typically~$\simeq
4$). We run the test-particle code and RMHD code at the same time so
that we can update the field information for the particles as often as
needed without saving the full time history of the RMHD fields to
memory.  To save computational resources, we use a single RMHD
simulation to simulate, simultaneously, several different cohorts of
particles, where each cohort corresponds to a different choice of the
parameters $\rho_{\rm i}/L_\perp$, $\epsilon$, and $\beta_{\rm i}$.
In particular, each simulation designation in Table~\ref{tab:summary}
(e.g., D1, D2, etc.) corresponds to several different test-particle
cohorts (each with a different value of~$\epsilon$) within the same
RMHD simulation.

We define~$N$ to be the number of particles in each cohort of
particles.  We estimate the numerical error associated with finite
particle number~$N$ by measuring the difference between the heating
rate determined using $N$ particles and the heating rate determined
using $N/2$ particles. We find that this error is $\simeq 5\%$ for $N
= 5.12 \times 10^4$. We use $N\simeq 10^5$ in the simulation results
reported in Section~\ref{sec:results}.

\subsection{Charge-to-Mass Ratios, Physical Lengthscales, and the RMHD Assumptions}
\label{sec:physical} 

The test particles in our simulations can represent any ion
species. Fundamentally, this is because physical lengthscales such as
the proton gyroradius and proton inertial length do not enter into the
RMHD equations. There are thus several ``hidden'' parameters whose
values we are free to adjust in order to make the test particles
correspond to protons, alpha particles, or a minor-ion species. For
example, after fixing the values of the dimensionless quantities
$\rho_{\rm i}/L_\perp$, $\epsilon$, and $\beta_{\rm i}$ as described
at the end of Section~\ref{sec:particle}, we can assign an arbitrary
physical value to $\delta u$, which translates into physical values
for $v_{\perp \rm i}$ and~$v_{\rm A}$ through our choices
of~$\epsilon$ and~$\beta_{\rm i}$. We are free to assign any physical
value to the proton mass density~$n_{\rm p} m_{\rm p}$; the values
of~$n_{\rm p} m_{\rm p}$ and~$v_{\rm A}$ then yield the value of~$B_0$
through Equation~(\ref{eq:defvA}).  We can then assign any physical
value to~$L_\perp$. Since the values of~$v_{\perp \rm i}$ and
$\rho_{\rm i}/L_\perp$ have already been specified, the value of
$L_\perp$ determines the physical value of~$\Omega_{\rm i}$.
Combining~$\Omega_{\rm i}$ and~$B_0$ yields the charge-to-mass
ratio~$q/m$, which can take on any value that we choose. In addition,
because we can separately assign arbitrary physical values to
both~$\delta u$ and~$L_\perp$, we can choose the the test-particle
ions to be any species we like and simultaneously take the inverse
proton gyroradius to correspond to any desired multiple of the
dissipation wavenumber~$k_{\rm d}$ of the turbulence power spectrum,
which is defined in Equation~(\ref{eq:defkd}) below.

If we take~$k_{\rm d}$ to correspond to the spectral break at
wavenumbers~$\sim \rho_{\rm p}^{-1}$ in the magnetic power spectrum in
the solar wind, then the perpendicular box size in the largest of our
simulations (Simulations D1 through~D5 of Table~\ref{tab:summary}) is
150 times larger than the wavelength $2\pi /k_{\rm d}$ corresponding
to~$k_{\rm d}$. Because the inertial range spans more than three
orders of magnitude in wavenumber in the solar wind at $r\lesssim 1
\mbox{ AU}$, our RMHD simulations approximate just a portion of the
inertial range --- the largest $\sim 2$ orders of magnitude of
wavenumbers just below~$k_{\rm d}$. 

As mentioned at the beginning of Section~\ref{sec:RMHD}, one of the
assumptions of RMHD is that the rms amplitude of the magnetic-field
fluctuations $\delta B_{ \rm rms}$ satisfies
\begin{equation}
\delta B_{\rm rms} \ll B_0.
\label{eq:rmhdlimit} 
\end{equation} 
We can write the ratio $\delta B_{\rm rms}/B_0$ in the form
\begin{equation}
\frac{\delta B_{\rm rms}}{B_0}  = \left(\frac{\delta B_{\rm rms}
    v_{\rm A}}{B_0 u_{\rm rms}}\right) \left(\frac{u_{\rm rms}}{\delta
  u}\right) \epsilon \beta_{\rm i}^{1/2}.
\label{eq:dbrmhd} 
\end{equation} 
The quantities $\delta B_{\rm rms}$ and $u_{\rm rms}$ are the total
rms values of the fluctuating magnetic field and velocity,
respectively, including contributions from all values of~$k_\perp$.
The first term on the right-hand side of Equation~(\ref{eq:dbrmhd}),
$\delta B_{\rm rms} v_{\rm A}/(B_0 u_{\rm rms})$, is $\simeq 1.4$ in
our simulations. The second term on the right-hand side of
Equation~(\ref{eq:dbrmhd}), $u_{\rm rms}/\delta u$, is $\simeq 2.2$.  In our
simulations with $\beta_{\rm i} < 0.2$, we restrict $\epsilon$ to
values~$\lesssim 0.25$.  This restriction leads to small values of
$\delta B_{\rm rms}/B0$ in our simulations with $\beta_{\rm i} =
0.006$, specifically,
\begin{equation}
\frac{\delta B_{\rm rms} }{B_0} < 0.07 \hspace{0.3cm} \mbox{(at $\beta_{\rm i} = 0.006$)}.
\label{eq:dBmax_lowbeta} 
\end{equation} 
In our larger-$\beta_{\rm i}$ simulations, on the other hand,
Equation~(\ref{eq:rmhdlimit})  is less well satisfied. The most
problematic case is our~$\beta_{\rm i} =1$ simulations, in which
we restrict $\epsilon$ to
values smaller than~$0.157$, leading to a maximum value of $\delta
B_{\rm rms}/B_0$ of~0.47, so that
\begin{equation}
\frac{\delta B_{\rm rms} }{B_0} \leq 0.47 \hspace{0.3cm} \mbox{ (at $\beta_{\rm i} = 1$)}.
\label{eq:dBmax_highbeta} 
\end{equation} 
We return to this issue when we discuss Figures~\ref{fig:qprap}
and~\ref{fig:xia_c1c2_beta} below.

Another assumption in RMHD that was mentioned at the beginning of
Section~\ref{sec:RMHD} is that the lengthscales are much larger
than~$\rho_{\rm p}$. The application of our simulations to the
stochastic heating of thermal protons is thus approximate at best, and
some caution must be exercised when making inferences about stochastic
proton heating based on this work.  The use of RMHD simulations is
more justified for heavy ions whose gyroradii exceed~$\rho_{\rm p}$.
To treat the stochastic heating of thermal protons more rigorously, a
numerical approach that accounts for kinetic processes at lengthscales
$\sim \rho_{\rm p}$ is needed --- e.g., gyrokinetic simulations,
hybrid simulations, or particle-in-cell simulations. An advantage of
RMHD, and one of the reasons we use it here, is that it is possible to
simulate a large dynamic range in three dimensions in RMHD while
simultaneously focusing computational resources on highly anisotropic
fluctuations with $k_\perp \gg k_\parallel$. This large dynamic range
makes it possible to explore phenomena such as coherent structures in
3D turbulence, which become increasingly prominent as the inertial
range broadens~\citep{wan12b}. We return to this point in
Section~\ref{sec:results}.

\vspace{0.2cm}
\section{Results}
\label{sec:results}
\vspace{0.2cm}

We have carried out RMHD simulations for four different grid sizes:
$128^3$, $256^3$, $512^2 \times 256$, and $1024^2\times 256$. These
grid sizes are listed in the format $N_\perp^2 \times N_\parallel$,
where $N_\parallel$ is the number of grid points along the direction
of the background magnetic field (the $z$ direction), and $N_\perp$ is
the number of grid points in each of the $x$ and $y$ directions. 
As $N_\perp$ increases, we decrease
the viscosity~$\nu$ so that the Reynolds number~Re increases, where
\begin{equation}
\mbox{Re} = \frac{u_{\rm rms} L_\perp}{\nu}.
\label{eq:defRe} 
\end{equation} 
We inject equal amounts of energy into $z^+$ and $z^-$ so
that the turbulence is ``balanced.'' 
For reference, we  define a dissipation wavenumber
\begin{equation}
k_{\rm d} = \left[\frac{\displaystyle\int k _\perp ^4 E_u(k_\perp)dk_\perp}{\displaystyle\int k_\perp ^2 E_u(k_\perp)dk_\perp}\right]^{1/2}.
\label{eq:defkd} 
\end{equation}
The values of $k_{\rm d}$ and Re in our different RMHD
simulations are listed in Table~\ref{tab:summary}. As mentioned in
Section~\ref{sec:particle}, we track several different cohorts of
particles within each RMHD simulation, where each cohort has a
different value of~$\epsilon$ but the same value of~$\beta_{\rm i}$. 

\begin{figure}[!htp]
\includegraphics[width=8.2cm]{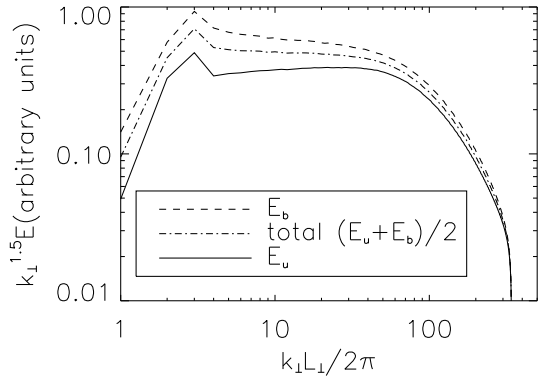}
\caption{The velocity power spectrum $E_u$, magnetic power spectrum
  $E_b$, and total-energy spectrum $(E_u + E_b)/2$ averaged over two
  of our $1024^2 \times 256$ simulations (D1 and~D2).
 \label{fig:vbspe}}
\end{figure}

At $k_\perp \gtrsim k_{\rm d}$, the fluctuations are strongly
influenced by dissipation.  At much smaller wavenumbers $\sim 2\pi
/L_\perp$, the fluctuations are strongly influenced by the details of
the numerical forcing term~$\bm{f}^\pm$ in
Equation~(\ref{eq:ch1incompress}).  However, in the ``inertial range''
of wavenumbers satisfying $(2\pi/L_\perp) \ll k_\perp \ll k_{\rm d}$,
forcing and dissipation have only a small effect, and the power
spectra attain approximately power-law forms.
 The velocity power spectrum
$E_u$, magnetic power spectrum $E_b$, and total power spectrum $E_{\rm
  tot} = (E_u + E_b)/2$, averaged over two of 
 our $1024^2 \times 256$ simulations (D1 and D2), are
shown in Figure~\ref{fig:vbspe}. As this figure shows, 
in the inertial range
 $E_{\rm tot} \propto k^{-3/2}$, $E_u$ is somewhat
flatter than $k^{-3/2}$ and $E_b$ is somewhat steeper than
$k^{-3/2}$, as in previously published RMHD
simulations~\citep{boldyrev11a}.

\begin{figure}[!htp]
\includegraphics[width=8.cm]{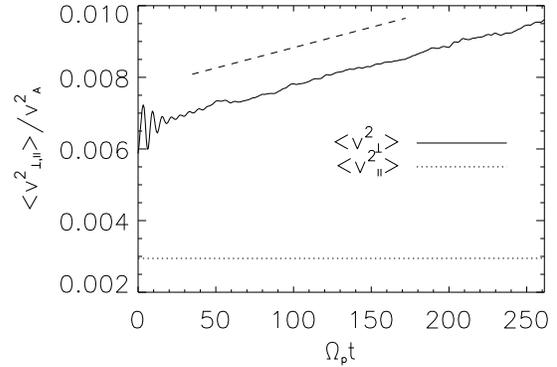}
\caption{ $v^2 _\perp$ and $v^2 _\parallel$ vs time for a particle
  cohort with $\epsilon = 0.15$ in Simulation~D1. The dashed line
  indicates the fitting range we use for finding
  $Q_\perp$. \label{fig:Q_vs_t} }
\end{figure}

To calculate $Q_\perp$ in our simulations, we measure the mean square
values of the particles' velocity components perpendicular and
parallel to the local magnetic field, $v_\perp$ and $v_\parallel$.  In
Figure~\ref{fig:Q_vs_t}, we plot $\langle v_\perp^2\rangle$ and
$\langle v_\parallel^2\rangle$ versus time in a $1024^2 \times 256$
simulation with $\rho_{\rm i} = L_\perp/(20\pi)$, $\epsilon=0.15$, and
$\beta_{\rm i} = 0.006$. (This figure describes one of the particle
cohorts in Simulation~D1 of Table~\ref{tab:summary}.)  This value
of~$\beta_{\rm i}$ (and all values of~$\beta_{\rm i}$ listed in
Table~\ref{tab:summary}) describe the initial particle
distribution. During the first few gyroperiods after the particles are
introduced into the simulation, $\langle v_\perp^2\rangle$ increases
rapidly as the particles ``pick up'' the $\bm{E}\times\bm{B}$ velocity
of the turbulence.\footnote{In the limit of small~$\rho_{\rm i}$, the
  particles would pick up the full $\bm{E}\times\bm{B}$
  velocity. However, some of the turbulent fluctuations are on scales
  $\lesssim \rho_{\rm i}$, and the interaction between the particles
  and these small-scale fluctuations is more complicated than a simple
  $\bm{E}\times\bm{B}$ drift.} We neglect this brief period of
transient heating when calculating the heating rates. More
specifically, we set
      \begin{equation}
      	Q_\perp = \frac{1}{2} \left( \frac{\langle v_{\perp \rm f} ^2\rangle - \langle v_{\perp 0} ^2\rangle}{t_{\rm f} - t_0} \right) \label{eq:ch4qave},
      \end{equation}
and 
      \begin{equation}
      	Q_\parallel = \frac{1}{2} \left( \frac{\langle v_{\parallel \rm f} ^2\rangle - \langle v_{\parallel 0} ^2\rangle}{t_{\rm f} - t_0} \right) \label{eq:ch4q_par_ave},
      \end{equation}
where $\langle ... \rangle$ indicates an average over all simulated
particles, $v_{\perp 0}$ ($v_{\perp \rm f}$) is a particle's
perpendicular velocity at $t = t_0$ ($t = t_{\rm f}$), $t _0 = 10 /
\Omega_{\rm i}$, and $t_f$ is either the end of the simulation or the time
at which $\langle v_{\perp f} ^2\rangle = 1.2 \langle v_{\perp 0}
^2\rangle$. The reason we do not use values of $t_{\rm f}$ that are so
large that $\langle v_{\perp f} ^2\rangle > 1.2 \langle v_{\perp 0}
^2\rangle$ is that the heating rate decreases as $\langle v_{\perp \rm
  f}^2 \rangle $ increases~\citep{chandran10}.

\begin{figure}[!htp]
\includegraphics[width=8.cm]{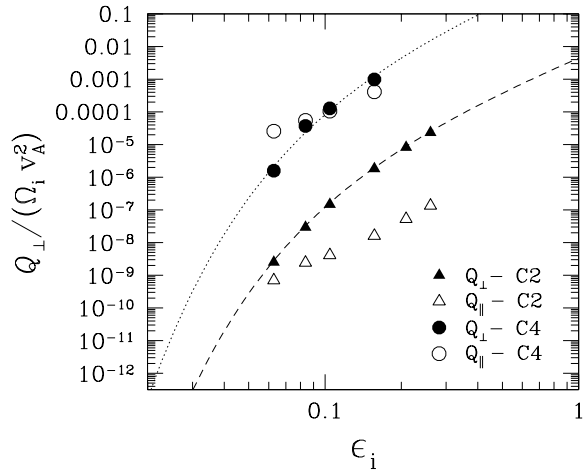}
\vspace{-1cm}  
\caption{$Q_\perp$ and $Q_\parallel$ as functions of $\epsilon_{\rm
    i}$ in Simulation~C2 (in which $\beta_{\rm i} = 0.006$) and
  Simulation~C4 (in which $\beta_{\rm i} = 1$). The dashed line is a
  plot of Equation~(\ref{eq:Q2}) with $c_1 = 1.1$ and $c_2= 0.4$, and
  the dotted line is a plot of Equation~(\ref{eq:Q2}) with $c_1 =
  3.7$ and $c_2 = 0.38$. This plot shows that $Q_\parallel/Q_\perp$
  increases as $\beta_{\rm i}$ increases.
\label{fig:qprap}}
\end{figure}

For each value of $\rho_{\rm i}/L_\perp$, $\beta_{\rm i}$, and
$N_\perp^2 \times N_\parallel$ that we investigate, we simulate
$N_{\rm c}$ cohorts of particles with different initial values
of~$\epsilon$. When $\beta_{\rm i} < 0.2$, we use $N_{\rm c} =
6$. When $\beta_{\rm i} > 0.2$, we use $N_{\rm c} = 4$ (because we
restrict ourselves to a smaller maximum value of~$\epsilon$ in order
to reduce the maximum value of $\delta B_{\rm rms}/B_0$, as discussed
in Section~\ref{sec:physical}). For each cohort, we calculate
$Q_\perp$ using Equation~(\ref{eq:ch4qave}). We then fit these values
of $Q_\perp$ to Equation~(\ref{eq:Q2}), varying $c_1$ and $c_2$ to
optimize the fit.  (Figure~\ref{fig:qprap} illustrates two such fits.)
We thus obtain a value of $c_1$ and $c_2$ for each choice of
$\rho_{\rm i}/L_\perp$, $\beta_{\rm i}$ and $N_\perp^2\times
N_\parallel$.  We list the results of our simulations in
Table~\ref{tab:summary}, along with several important parameters,
including
\begin{equation}
k_\rho = \frac{1}{\rho_{\rm i}}.
\label{eq:defkrho} 
\end{equation} 
It is fluctuations with $e^{-0.5} k_\rho < k_\perp < e^{0.5} k_\rho$
that determine the value of $\delta u$ in Equation~(\ref{eq:defvp})
and hence the value of~$\epsilon$ in Equations~(\ref{eq:Q2}) and
(\ref{eq:defeps}). Because $\langle v_\perp^2\rangle$ and~$\delta u$
vary during a simulation, the value of~$\epsilon$ also varies. When we
use our numerical simulations to determine~$Q_\perp$ and~$Q_\parallel$
as functions of~$\epsilon$, we calculate $\epsilon$ in
Equation~(\ref{eq:defeps}) using the value of~$v_{\perp \rm i}$ at the
instant the particles are introduced and the value of~$\delta u$ that
is obtained by averaging the velocity power spectrum over the time
interval~$(t_0, t_{\rm f})$ during which the heating rates are
calculated. Thus, the final values of~$\epsilon$ that are used in,
e.g., Figure~\ref{fig:qprap} below, differ slightly from the values
of~$\epsilon$ at the beginning of the test-particle integration.

\begin{table}[!h]
\caption{Simulation Parameters}
\begin{center}
\begin{tabular}{ c c c c c c c c}
\hline
\hline
\vspace{-0.3cm} 
\\
Run  & Grid Size & Re  & $\displaystyle \frac{k_{\rm d} L_\perp}{2\pi}$ & 
$\displaystyle \frac{k_{\rho} L_\perp}{2\pi}$ & $\beta_{\rm i}$ & $c_2$ & $c_1$
\vspace{0.1cm} 
\\ \hline
A  & $128^3$  	  	  &2400	&29  & 10  & 0.006 & 0.44 & 0.83\\
B1  & $256^3$  		  &6000	&42  & 10  & 0.006 & 0.41 & 1.0 \\

B2  & $256^3$  		  &6000	&42  & 10  & 0.033 & 0.41 & 1.5 \\
B3  & $256^3$  		  &6000	&42  & 10  & 0.18 & 0.42 & 1.6 \\
B4  & $256^3$  		  &6000	&42  & 10  & 1.0 & 0.41 & 3.6 \\

C1  & $512^2\times 256$   &15000	&77  & 10  & 0.006 & 0.29 & 0.80\\
C2  & $512^2\times 256$   &15000	&77  & 20  & 0.006 & 0.40 & 1.1\\

C3  & $512^2\times 256$   &15000	&77  & 20  & 0.1 & 0.37 & 0.86\\
C4  & $512^2\times 256$   &15000	&77  & 20  & 1.0 & 0.38 & 3.7\\

D1  & $1024^2\times 256$  &38000	&150 & 10   & 0.006 & 0.20 & 0.71\\
D2  & $1024^2\times 256$  &38000	&150 & 20   & 0.006 & 0.22 & 0.67\\
D3  & $1024^2\times 256$  &38000	&150 & 40   & 0.006 & 0.25 & 0.62\\
D4  & $1024^2\times 256$  &38000	&150 & 80   & 0.006 & 0.21 & 0.74\\
D5  & $1024^2\times 256$  &38000	&150 & 160  & 0.006 & 0.15 & 0.94\\
 \hline\hline
 \label{tab:summary}
\end{tabular}
\end{center}
\end{table}

As Table~\ref{tab:summary} shows, both $c_1$ and $c_2$ vary from
simulation to simulation. However, because $c_2$ appears in the
argument of the exponential function in Equation~(\ref{eq:Q2}), the
variations in the values of $c_2$ are particularly important 
for modeling stochastic heating in the solar
corona and solar wind. Much of our focus is thus on how~$c_2$ depends upon
simulation size, $k_\rho$, and $\beta_{\rm i}$. 

\begin{figure}[!htp]
\includegraphics[width=7.cm]{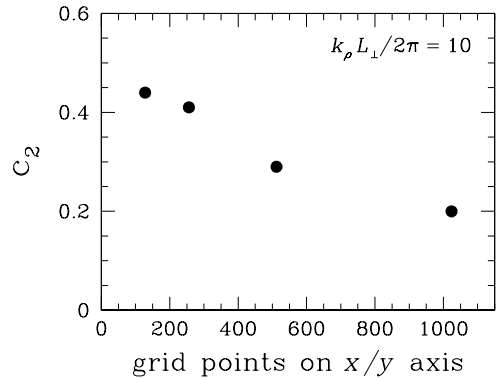}
\vspace{-1cm}  
\caption{The dependence of $c_2$ on grid resolution (and Re --- see
  Table~\ref{tab:summary}) at fixed $k_\rho =
  10$. The data points are from Simulations A, B1, C1, and D1 in
  Table~\ref{tab:summary}. In these simulations, $\beta_{\rm i} =
  0.006$. \label{fig:fixedkrho} }
\end{figure}

In Figure~\ref{fig:fixedkrho} we plot the values of $c_2$ in
simulations A, B1, C1, and~D1 of Table~\ref{tab:summary}. These
simulations have the same values of $k_\rho L_\perp/(2\pi)$ and
$\beta_{\rm i}$ (10 and 0.006, respectively), but different values
of~$N_\perp$ and~$\mbox{Re}$. This figure shows that $c_2$ decreases
(i.e., stochastic heating becomes stronger) at fixed $k_\rho L_\perp$
as $N_\perp$ and~Re increase. As $N_\perp$ and~Re increase at fixed
$k_\rho L_\perp$, the inertial range extends to larger wavenumbers
relative to~$k_\rho$ (i.e., $k_{\rm d}/k_\rho$ increases), and there
is more wave power at $k_\perp \gtrsim k_\rho$ for any fixed value
of~$\epsilon$.  This additional small-scale power enhances the
stochastic heating rate, providing an explanation for the trend in
Figure~\ref{fig:fixedkrho}.

\begin{figure}[!htp]
\includegraphics[width=7.cm]{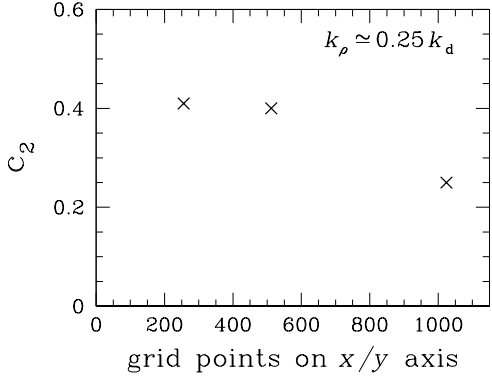}
\vspace{-1cm}  
\caption{The dependence of $c_2$ on grid resolution (and Re --- see
  Table~\ref{tab:summary}) at fixed
  $k_\rho/k_{\rm d} \simeq 0.25$. The data points are from Simulations
  B1, C2, and D3 in Table~\ref{tab:summary}. In these simulations,
  $\beta_{\rm i} = 0.006$. \label{fig:fixedkrho_over_kd} }
\end{figure}

\begin{figure}[!htp]
\includegraphics[width=8.cm]{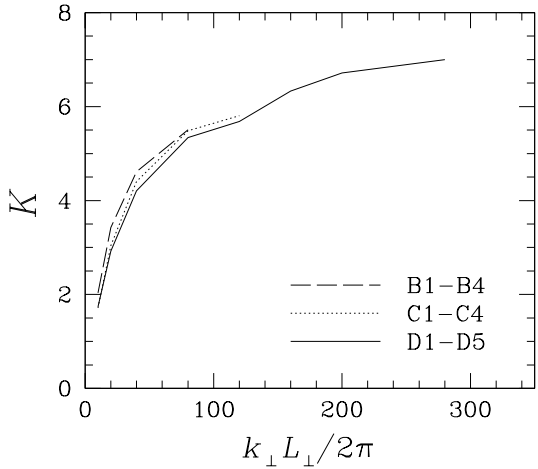}
\vspace{-1cm}  
\caption{Kurtosis $K$ as a function of~$k_\perp$ in Simulations B1-B4,
  C1-C4, and D1-D5. 
\label{fig:7kurtosis}}
\end{figure}

In Figure~\ref{fig:fixedkrho_over_kd}, we plot the values of $c_2$ in
Simulations B1, C2, and D3. The value of $N_\perp$ doubles going from
Simulation~B1 to Simulation~C2, and doubles again going from
Simulation~C2 to Simulation~D3. As $N_\perp$ doubles, the Reynolds
number is increased in such a way that the value of $k_d L_\perp$ also
approximately doubles in these simulations. For the simulations shown
in Figure~\ref{fig:fixedkrho_over_kd}, we double $k_\rho L_\perp$ each
time $N_\perp$ doubles, so that $k_\rho/k_{\rm d}$ remains
approximately constant. As this figure shows, doubling $k_\rho
L_\perp$ while keeping $k_\rho/k_{\rm d}$ constant causes $c_2$ to
decrease, so that stochastic heating becomes more effective.  There
are likely two reasons for this. First, as $k_\rho L_\perp$ increases
at fixed $k_\rho/k_{\rm d}$, there is extra fluctuation energy at
$k_\perp < k_\rho$ in the RMHD simulation, and this additional
fluctuation energy contributes to some extent to particle
heating. Second, as $k_\rho L_\perp$ increases, coherent structures
become more prevalent at the ion-gyroradius scale. Coherent
structures, such as current sheets, become increasingly prevalent at
smaller scales within the inertial range~\citep{wan12b}.  This
phenomenon can be seen in Figure~\ref{fig:7kurtosis}, where we plot
the kurtosis $K$ as a function of $k_\perp$ in Simulations~B1-B4, C1-C4,
and~D1-D5. The kurtosis is a measure of coherent structures or
non-Gaussianity.  For a distribution of $n$ vectors $\bm{x}_j$ with
$j=0,1,2,\dots, n-1$, the kurtosis is given by
\begin{equation}
K = \frac{1}{n} \sum_{j=0}^{n-1} \left| \frac{\bm{x}_j - \bar{\bm{x}}}{\sigma} \right|^4 - 3 ,
\label{eq:defK} 
\end{equation}
where $\bar{\bm{x}} = n^{-1} \sum_{j=0}^{n-1} \bm{x}_j$ and $ \sigma =
[(n-1)^{-1} \sum_{i=0}^{n-1} |\bm{x}_i - \bar{\bm{x}}|^2]^{1/2}.$ In
this problem, the individual vectors~$\bm{x}_j$ are of the form
$\bm{x}_j = \bm{z}^\pm(\bm{r}+\bm{l})-\bm{z}^\pm(\bm{r})$, where the
$\bm{r}$ vectors locate a set of evenly spaced grid points (separated
by eight grid points in each Cartesian direction), the separation
vector~$\bm{l}$ is given by either $\bm{l} = \bm{\hat{x}}\pi/k_\perp$
or $\bm{l} = \bm{\hat{y}} \pi/k_\perp$, and~$k_\perp$ is varied.  In
our strong RMHD turbulence simulations, $K$ increases steadily with
increasing~$k_\perp$, reaching values $\simeq 5$ at $k_\perp
L_\perp/2\pi \simeq 70$. The importance of coherent structures in
turbulent heating has been discussed by a number of authors
\citep[e.g.,][]{dmitruk04,wan10,parashar11,markovskii11,servidio11b,servidio12,
  greco12, tenbarge13,haynes13,karimabadi13,wu13}.  The reason that
coherent structures enhance stochastic heating was described by
\cite{chandran10}. As the turbulent heating is concentrated into a
smaller volume in which the fluctuation amplitudes are larger, the
particle orbits become more stochastic at the locations where the
heating occurs.

\begin{figure}[!htp]
\includegraphics[width=7.cm]{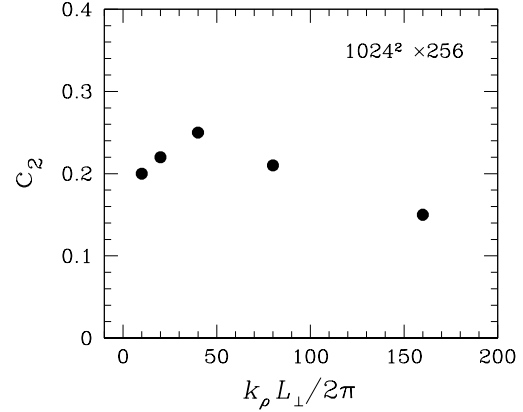}
\vspace{-1cm}  
\caption{The value of $c_2$ in Simulations~D1 through~D5, in
  which~$\beta_{\rm i} = 0.006$. The gyroradius $\rho_{\rm i} =
  k_\rho^{-1}$ takes on a different value in each simulation.
\label{fig:c2_1024} }
\end{figure}

In Figure~\ref{fig:c2_1024}, we show how $c_2$ depends upon the value
of $k_\rho$ in our~$1024^2 \times 256$ simulations. The data points in this
figure correspond to Simulations D1 through~D5 in
Table~\ref{tab:summary}. This figure appears to show a
competition between the effects described above. As $k_\rho
L_\perp/(2\pi)$ increases from $10$ to $40$, there are fewer modes
with $k_\perp$ exceeding~$k_\rho$ within the RMHD simulations, and thus
fewer sub-gyroradius-scale fluctuations contribute to the stochastic
heating of the test particles. We conjecture that this is why $c_2$
increases in Figure~\ref{fig:c2_1024} as $k_\rho L_\perp/(2\pi)$
increases from $10$ to $40$. On the other hand, as $k_\rho L_\perp/
(2\pi)$ increases above~40, the trend reverses and $c_2$ decreases.
We conjecture that this decrease occurs because two other effects
discussed above become dominant: the increasing prevalence of coherent
structures at $k_\perp = k_\rho$ and the increasing contribution to
$Q_\perp$ from fluctuations with $k_\perp \ll k_\rho$. This latter
effect becomes increasingly important as $k_\rho$ increases to values
$\gtrsim k_{\rm d}$. In this case, $\delta u$ becomes so small that
even a modest contribution to $Q_\perp$ from fluctuations with
$k_\perp \ll k_\rho$ leads to a significant decrease in~$c_2$ and/or
increase in~$c_1$.

Figure~\ref{fig:c2_1024} implies that there is some variation in the
value of~$c_2$ for different ion species. For example, if ion
temperatures are mass proportional, then $k_{\rho} \propto q/m$. In
this case, if the point with $k_\rho L_\perp /2\pi = 80$ in
Figure~\ref{fig:c2_1024} represents protons, then the point with
$k_\rho L_\perp /2\pi = 40$ corresponds to alpha particles. However,
some caution is warranted when attempting to infer the species
dependence of~$c_2$ from Figure~\ref{fig:c2_1024}. As discussed in
Sections~\ref{sec:physical} and~\ref{sec:cons}, our RMHD simulations
neglect kinetic effects that arise at scales~$\lesssim \rho_{\rm p}$,
which presumably have a larger effect on the value of~$c_2$ for
protons than on the value of~$c_2$ for heavy ions with gyroradii
exceeding~$\rho_{\rm p}$.

Figure~\ref{fig:qprap} shows the relative strength of parallel heating
and perpendicular heating as a function of~$\epsilon$ for two
different values of~$\beta_{\rm i}$. Because ideal RMHD fluctuations
possess a magnetic-field-strength fluctuation that is second order in
the fluctuation amplitude~($\delta |B|/B_0 \sim (|\delta
\bm{B}|/B_0)^2$) and have no parallel electric field,
Landau/transit-time damping (LD/TTD) is weaker than it would be in,
e.g., fast magnetosonic turbulence with the same rms value of~$|\delta
\bm{B}|/B_0$. Nevertheless, as these figures show, some parallel
heating does occur. Figure~\ref{fig:qprap} shows that
$Q_\parallel/Q_\perp$ is larger when $\beta_{\rm i} \sim 1$ than
when~$\beta_{\rm i} \sim 0.006$. This is in part because
when~$\beta_{\rm i} \ll 1$, the ions are too slow to ``surf''
effectively on the RMHD fluctuations, even accounting for the
resonance broadening that arises from the nonlinear decorrelation of
the fluctuations~\citep{lehe09,lynn12}. We note that our simulation
method does not provide a realistic assessment of parallel heating by
low-frequency Alfv\'en-wave turbulence in the solar wind, because such
turbulence likely becomes KAW turbulence at perpendicular scales
$\lesssim \rho_{\rm p}$. KAW fluctuations involve
magnetic-field-strength fluctuations that are first order in the
fluctuation amplitude ~($\delta |B|/B_0 \sim |\delta
\bm{B}|/B_0$). Parallel heating from LD/TTD is thus much stronger for
KAW turbulence than it is for RMHD turbulence.

As mentioned in Section~\ref{sec:physical}, when $\beta_{\rm i} =1$,
we restrict our simulations to $\epsilon \lesssim 0.15$ in order to
reduce the maximum value of~$\delta B_{\rm rms}/B_0$.  This is why the
results from Simulation~C4 in Figure~\ref{fig:qprap} are limited to
smaller $\epsilon$ values than the results from Simulation~C2.
Despite this restriction, the value of $\delta B_{\rm rms}/B_0$
reaches a maximum value of 0.47 in the $\beta_{\rm i} = 1$ simulation
with $\epsilon \simeq 0.15$. This value of~$\delta B_{\rm rms}/B_0$ is
not~$\ll 1$, as required in~RMHD, and thus our~$\beta_{\rm i} = 1$
results must be viewed with some caution. For example, because $\delta
B_{\rm rms}/B_0 \simeq 1/2$ in our simulation with $\beta_{\rm i} = 1$
and $\epsilon_{\rm i} \simeq 0.15$, the magnetic field lines in this
simulation are tilted by~$\simeq 30^\circ$ with respect to the~$z$
direction. The small-scale structure in the $xy$-plane in this
simulation may thus lead to small-scale structure parallel to the
local magnetic field lines \citep[but see][]{cho00}.  As a
consequence, some of the perpendicular heating in this simulation may
result from a Doppler-shifted cyclotron resonance. Alternative
simulation techniques, including test-particle simulations based on
incompressible MHD rather than RMHD, would be useful for further
investigations into stochastic-heating at~$\beta_{\rm i} = 1$.

\begin{figure}[!htp]
\includegraphics[width=8.cm]{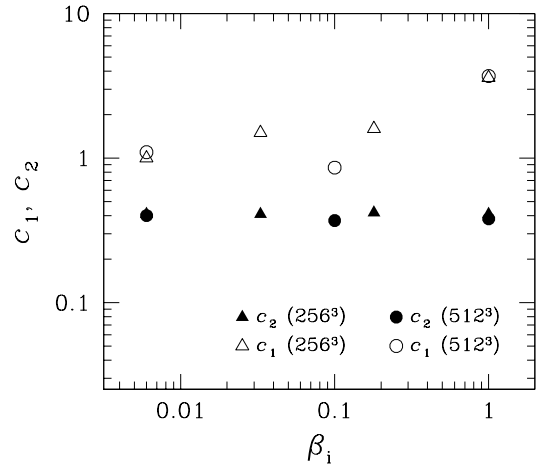}
\vspace{-1cm}  
\caption{$c_1$ and $c_2$ vs $\beta_{\rm i}$ in Simulations B1 to B4
  (triangles) and Simulations~C2 to~C4 (circles).\label{fig:xia_c1c2_beta}}
\end{figure}

In Figure~\ref{fig:xia_c1c2_beta} we plot the values of $c_1$ and
$c_2$ in several simulations with different values of
$\beta_{\rm i}$.  As $\beta_{\rm i}$ increases from 0.006 to~1, $c_2$
undergoes only small variations, but $c_1$ increases by a factor
of~$\sim 3$. These trends indicate that stochastic heating becomes 
more effective at fixed $\delta u$, $\rho_{\rm i}$, and~$\epsilon_{\rm
  i}$ as $\beta_{\rm i}$ is increased from~0.006 to~1.  However, as
discussed above, our results on the~$\beta_{\rm i} =1$ case must be
viewed with caution, since~$\delta B_{\rm rms}/B_0$ reaches values as
large as~0.47.

\vspace{0.2cm}
\section{Discussion and Conclusion}
\label{sec:cons}
\vspace{0.2cm}

In this paper, we describe numerical simulations of test-particles
interacting with strong RMHD turbulence. By tracking the change in
particle energy with time, we evaluate the stochastic heating
rate~$Q_\perp$ as a function of the amplitude of the turbulent
fluctuations at the gyroradius scale.  Our results for~$Q_\perp$ are
well described by the functional form in Equation~(\ref{eq:Q2}), which
was derived by \cite{chandran10} using phenomenological arguments. Our
simulations enable us to evaluate the constants~$c_1$ and $c_2$ in
this equation for strong RMHD turbulence, and to determine how these
constants depend upon various properties of the turbulence and
particle distributions.

We find that strong RMHD turbulence is much more effective than
randomly phased waves at stochastically heating ions, in the sense
that the constant $c_2$ in Equation~(\ref{eq:Q2}) is significantly
smaller for ions interacting with strong RMHD turbulence than for ions
interacting with randomly phased waves. This difference likely arises
because of the coherent structures that develop in strong RMHD
turbulence, whose prevalence is measured by the kurtosis~$K$ in
Equation~(\ref{eq:defK}).  The fluctuation amplitudes are larger in
the vicinity of coherent structures, which increases orbit
stochasticity. This enables particles to absorb more energy from the
time-varying electrostatic potential in the regions where most of the
particle heating occurs.  We also find that the constants~$c_1$ and
$c_2$ undergo small variations as $\beta_{\rm i}$ increases from 0.006
to~1, in the sense that stochastic heating becomes moderately more
effective for fixed values of~$\delta u$, $\rho_{\rm i}$,
and~$\epsilon_{\rm i}$.  This implies that stochastic heating can
occur not only in the low-$\beta_{\rm i}$ conditions of the solar
corona but also in the $\beta_{\rm i} \sim 1$ conditions found in the
solar wind near Earth.  Furthermore, we find that the parallel proton
heating rate~$Q_\parallel$ is much smaller than~$Q_\perp$ when
$\beta_{\rm i} \ll 1$ for the range of $\epsilon$ values that we have
investigated. On the other hand, $Q_\parallel$ can exceed~$Q_\perp$ in
our simulations for $\beta \sim 1$ and $\epsilon \lesssim 0.1$.  As
discussed in Section~\ref{sec:results}, our simulations underestimate
the amount of parallel heating in solar-wind turbulence, because we do
not take into account the change in the polarization properties of the
fluctuations at $k_\perp \rho_{\rm p} \simeq 1$, where the AW cascade
transitions into a KAW
cascade~\citep{bale05,howes08b,sahraoui09,schekochihin09}.

Our largest runs are Simulations D1 through~D5. The values of $c_2$ in
these simulations are $\sim 0.2$. On the other hand, the value of
$c_2$ steadily decreases as we increase the number of grid points and
Reynolds number, whether we hold constant $k_\rho L_\perp$ or
$k_\rho/k_{\rm d}$. If we were able to simulate RMHD turbulence with
an inertial range as broad as that found in the solar wind, then the
resulting values of $c_2$ would be smaller than the values that we
have found in Simulations~D1 through~D5. 

Previous studies have found that $c_2$ values of order or slightly
smaller than 0.2 are required in order for stochastic heating to
explain the observed ion temperatures in fast-solar-wind streams and
coronal holes. For example, \cite{bourouaine13} used {\em Helios}
magnetometer data to measure the amplitudes of the gyroscale
magnetic-field fluctuations in fast-solar-wind streams at heliocentric
distances between $0.29$ and~$0.64$~AU. Given these amplitudes, these
authors found that values of $c_2$ of $\simeq 0.2$ were sufficient for
stochastic heating to explain the non-adiabatic perpendicular proton
temperature profile measured by {\em Helios}.  These authors were
unable to determine the precise value of~$c_2$ that is required,
however, because of the uncertainty in the assumed relationship
between the amplitude of the gyroscale magnetic field fluctuation and
the amplitude of the gyroscale $\bm{E} \times \bm{B}$
velocity. \cite{chandran10b} used an approximate model of RMHD-like
turbulence in the extended solar atmosphere with $E_u \propto
k_\perp^{-3/2}$ and found that stochastic heating could explain
observations of $\mbox{O}^{+5}$ and proton temperatures in coronal
holes if $c_1 \simeq 1$ and $c_2 = 0.15$.  \cite{chandran11} assumed
that $c_1 = 0.75$ and $c_2 = 0.17$ in a two-fluid (proton/electron)
solar-wind model that incorporated Alfv\'en-wave turbulence,
stochastic heating, and proton temperature anisotropy and obtained a
reasonable match between the model perpendicular proton temperatures
and observations of coronal holes and fast solar-wind streams.

The approximate correspondence between
the values of $c_2$ in our $1024^2 \times 256$ simulations and the
values of~$c_2$ that are needed to explain observed ion temperatures
suggests that stochastic heating plays an
important role in the solar wind and coronal holes.
However, there are several reasons why the rate of stochastic
heating in the solar wind might differ from the rate in our numerical
simulations. First, we have only considered test particles and have
not accounted for the back reaction of the particles upon the
turbulence.  Second,
by considering RMHD turbulence, we neglect the kinetic-Alfv\'en-wave
(KAW) physics that arises at $k_\perp \rho_{\rm p} \gtrsim 1$. In KAW
turbulence, the electric-field power spectrum flattens as $k_\perp$
increases above~$\rho_{\rm p}^{-1}$~\citep{bale05}, and the additional
electric-field power at $k_\perp \rho_{\rm p} > 1$ presumably enhances
stochastic proton heating in AW/KAW turbulence above the level in RMHD
turbulence with the same value of~$\delta u$. Third, we have
neglected the effects of cross helicity or ``imbalance'' (which
arises, e.g., when more Alfv\'en waves propagate away from the Sun
than toward the Sun) and differential flow between ion species
\cite[but see][]{chandran13}. Future work, including numerical
simulations, that accounts for this
additional physics will be important for advancing our understanding
of stochastic heating further.

Finally, we note that a number of studies find that sheet-like
concentrations of current density play an important role in the
dissipation of solar-wind
turbulence~\citep[e.g.,][]{dmitruk04,servidio11b,wan12,karimabadi13}. Although
these structures are not singular, they are some times referred to as
``current sheets.''  Our results are consistent with these studies, in
the sense that we find that coherent structures enhance the stochastic
heating rate.

\acknowledgements We thank Peera Pongkitiwanichakul and the anonymous
referee for helpful suggestions and comments. This work was supported
in part by NSF grant AGS-0851005, NSF/DOE grant AGS-1003451, DOE grant
DE-FG02-07-ER46372, and NASA grants NNX07AP65G, NNX08AH52G, NNX11AJ37G
and NNX12AB27G. B.~Chandran was supported  by a Visiting
Research Fellowship from Merton College, University of Oxford.
E.~Quataert was supported by a Simons Investigator award from the
Simons Foundation, the David and Lucile Packard Foundation, and the
Thomas Alison Schneider Chair in Physics at UC Berkeley.

\bibliography{articles}

\end{document}